\documentclass[useAMS]{mn2e}

\usepackage{graphicx}
\def\approxgt{\mathrel{\hbox{\rlap{\lower.55ex \hbox {$\sim$}}
        \kern-.3em \raise.4ex \hbox{$>$}}}}
\def\approxlt{\mathrel{\hbox{\rlap{\lower.55ex \hbox {$\sim$}}
        \kern-.3em \raise.4ex \hbox{$<$}}}}

\title[Broad K$_{\alpha}$ Fe line in NGC~5506]{The final verdict by XMM-Newton: the X-ray obscured Seyfert galaxy NGC~5506 has a broad Fe K$_{\alpha}$ line}
\author[M.~Guainazzi et al.]{
M.~Guainazzi\thanks{E-mail:Matteo.Guainazzi@sciops.esa.int},$^{1}$
S.~Bianchi,$^{2}$
G.~Matt,$^{2}$
M.~Dadina,$^{3}$
J.~Kaastra,$^{4}$
J.~Malzac,$^{5}$
\newauthor
G.~Risaliti,$^{6,7}$ \\
$^{1}$European Space Astronomy Center of ESA, P.O.Box 78, Villanueva de la Ca\~nada, E-28691 Madrid, Spain \\
$^{2}$Dipartimento di Fisica, Universit\'a degli Studi Roma Tre, via della Vasca Navale 84, I-00046 Roma, Italy \\
$^{3}$INAF/IASF-Bo, via Gobetti 101, I-40129 Bologna, Italy \\
$^{4}$SRON Netherlands Institute for Space Research, Sorbonnelaan 2, 3584 CA Utrecht, The Netherland \\
$^{5}$Centre d'Etude Spatiale des Rayonnements, Universit\'e de Toulouse, CNRS, 9 avenue du Colonel Roche, BP 44346, 31028 Toulouse Cedex 4, \\ France \\
$^{6}$Harvard-Smithsonian Center for Astrophysics, 60 Garden St. Cambridge, MA 02138, USA \\
$^{7}$INAF-Osservatorio di Arcetri, Largo E. Fermi 5, I-50125, Firenze, Italy
}
\begin{document}

\date{}

\pagerange{\pageref{firstpage}--\pageref{lastpage}} \pubyear{2010}

\maketitle

\label{firstpage}

\begin{abstract}
We present the first unambiguous evidence of a broad (Gaussian width $\sim$330~eV) component of the
iron K$_{\alpha}$ fluorescent emission line in the X-ray obscured Narrow Line Seyfert~1
Galaxy NGC~5506. This is the main results of a spectroscopic monitoring campaign on this source
performed with the XMM-Newton
observatory between February 2001 and January 2009. The broad line lacks extreme redwards skewness.
If modelled with a relativistic component, the profile of the line is consistent with
a flat emissivity radial dependence ($\alpha \simeq$1.9).
The disk inclination ($\simeq$40$^{\circ}$) is nominally larger then typically observed in unobscured AGN,
in agreement with most measurements of broadened iron lines in Seyfert~2 galaxies. The quality of the
data allows us to decompose the full iron emission line complex, and
to study its long-term (timescales of weeks to years) variability pattern. The intensity of the
neutral and narrow iron K$_{\alpha}$ core remains  constant during the monitoring campaign. This indicates
that the optically thick gas responsible for the non-relativistic reprocessing of the primary AGN continuum
in NGC~5506 is probably located in the torus rather than in the optical Broad Line Region. 
\end{abstract}


\section{Introduction}

Broadening by relativistic effects of emission lines in an X-ray illuminated
accretion disk offers the opportunity of probing the physical conditions and geometrical distribution of matter in
the vicinity of black holes, as well as General Relativity effects \cite{reynolds03,miller07}.
Broad and skewed profiles of the iron K$_{\alpha}$ fluorescent emission line were discovered by ASCA
in the mid-90s
\cite{tanaka95,mushotzky95}. High-throughput observations with the EPIC cameras on-board XMM-Newton
have shown that at least 30 percent of nearby Seyfert galaxies exhibit relativistically blurred features consistent
with being produced within 50 gravitational radii ($r_g \equiv GM/c^2$, where $M$ is the black hole mass)
\cite{nandra07}. However, doubts have been cast
on the robustness and uniqueness of these results. Ionised outflows partially covering the primary
high-energy emission in AGN could mimic relativistic effects \cite{turner09,reeves04}.
Controversy is fierce. In this context, it is crucial to confirm whether objects, whose data had poor
signal-to-noise, do exhibit relativistically broadened lines
as well as to expand the parameter space covered by these measurements.

The study of X-ray obscured AGN is particularly promising. In the ``Unified Model''
framework \cite{antonucci85,antonucci93}, their putative accretion disks are expected to
be observed at higher inclination angles if the plane of the disk is aligned with that
of the obscuring matter. For inclinations not too close to ``edge-on'', special relativistic
effects, such as Doppler boosting, should be therefore stronger and easier to
detect. Moreover, in type~II objects the polar outflows should be out
of the line of sight and this should break the degeneracy between "warm
absorbers" and relativistic lines \cite{turner09}. Finally
relativistic lines allow the direct measurements of the inclination
angle of the accretion disk thus offering a simple test of the unified scenarios.

Observations of broad Fe line in X-ray obscured Seyferts are so far sparse. The Fe K$_{\alpha}$
line in the {\it Suzaku} spectrum of
MCG-5-23-16 implies a disk inclination, $\imath$, of
$53^{\circ}$$\pm^7_9$$^{\circ}$ \cite{reeves07}\footnote{see, however, Nandra et al.
(2007), whose spectral deconvolution of the XMM-Newton/EPIC spectra requires $\imath$$\approxlt$20$^{\circ}$}.
Nandra et al. (2007) add to this census NGC~526A ($\imath$=$43^{\circ}$$\pm^{42}_{20}$$^{\circ}$)
and NGC~2992 ($\imath$=$24^{\circ}$$\pm 7$$^{\circ}$), whereas
NGC~2110 and Mkn~6 do not exhibit significant evidence for relativistic effects.
NGC~1365 ($\imath$=$24^{\circ}$$\pm^8_4$$^{\circ}$; Risaliti et al. 2009) and
IRAS13197-1627 ($\imath$=$27^{\circ}$$\pm 17$$^{\circ}$; Dadina \& Cappi 2004, Miniutti et al. 2007b)
were also recently
discovered to host broad lines from nominally low inclination disks.
The aforementioned uncertainties are purely statistics; systematic uncertainties related {\it e.g.}
to different spectral deconvolutions are seldom discussed in the literature.
They can be at least comparable to the statistical ones.

In this context, the case
of the nearby ($z = 0.0061$) X-ray obscured ($N_H \simeq 3 \times 10^{22}$~cm$^{-2}$; Wang et al. 1999)
Narrow Line Seyfert~1
Galaxy (NLSy1; Nagar et al. 2002) NGC~5506 is one of the most controversial. Bianchi et al. (2003) analysed all the X-ray
observations performed with imaging and spectroscopic instruments on NGC~5506 at that time. They did not detect any
spectral components directly related to the accretion disk. This led them to conclude that the disk is either
fully ionised, or almost edge-on. Two early XMM-Newton observations of NGC~5506 (originally presented by
Matt et al. 2001) are also in the Nandra et al. (2007) sample. In their analysis, only one observation shows
evidence of broadened emission. Nandra et al. (2007) interpreted the lack of detection as simply due to
low signal-to-noise ratio.

In this paper we present spectroscopic results of a XMM-Newton monitoring campaign of NGC~5506, which
brings the total integration time to $\simeq$197~ks (versus the 23~ks of the observations published so far).
The main results of this paper is the {\it first unambiguous detection of a broad iron K$_{\alpha}$
line in NGC~5506}.

\section{Observations and data reductions}

Tab.~\ref{tab6} lists the XMM-Newton observations of NGC~5506 performed so far. We discuss in this
\begin{table}
  \caption{Log of XMM-Newton NGC~5506 observations}
  \label{tab6}
  \begin{tabular}{lcccc}
  \hline
  XMM-N~ID & Start & $\rho$$^a$ & $T_{CR}$$^b$ & T$_{exp}$$^c$ \\
          & Time & (\arcsec) & (s$^{-1}$) & (ks) \\
  \hline
  0013140101 (1)$^d$ & 02-02-2001 & 100/40 & 0.2/0.5 & 13.3 \\
  0013140201 (2)$^d$ & 09-01-2002 & 120/44 & 0.5/1.0 & 9.9 \\
  0201830201 (3)$^d$ & 11-07-2004 & 42/45 & 0.5/0.5 & 14.8 \\
  0201830301 (4)$^d$ & 14-07-2004 & 40/42 & 0.5/0.35 & 13.9 \\
  0201830401 (5)$^d$ & 22-07-2004 & 40/42 & 0.35/0.35 & 13.8 \\
  0201830501 (6)$^d$ & 07-08-2004 & 40/42 & 0.35/0.35 & 13.9 \\
  0554170101 (7)$^d$ & 27-07-2008 & 50/50 & 0.35/0.35 & 55.8 \\
  0554170201 (8)$^d$ & 02-01-2009 & 40/40 & 0.5/0.35 & 62.0 \\
 \hline
\end{tabular}

\noindent
$^a$size of the source spectrum extraction region in the MOS/pn;
$^b$count rate threshold applied to select intervals of low particle background
in the MOS/pn;
$^c$pn exposure time;
$^d$label used to identify the observation in this paper: XMM$i$
\end{table}
paper data of the EPIC cameras only
(MOS, Turner et al. 2001; pn, Str\"uder et al. 2001).
Data were reduced
using the SASv9.0 \cite{gabriel03} tasks {\tt e[mp]proc} with standard settings. Periods of high-background
were removed using count rate thresholds on the high-energy, single events, field-of-view light curves
(Tab.~\ref{tab6}) which optimise the signal-to-noise of each individual observation.
Source spectra were extracted from circular regions surrounding the source centroid (Tab.~\ref{tab6}),
whereas background spectra from circular regions on the same chip free from contaminating serendipitous sources
(and at the same height in detector coordinates for the pn to ensure that the same Charge Transfer Inefficiency
correction applies). Response files were generated with the SAS tasks {\tt arfgen} and {\tt rmfgen}.
Spectra were rebinned to over-sample the intrinsic energy resolution by a factor not
larger than 3, and to ensure that each spectral channel has a number of background-subtracted counts larger
than 50. Spectral fits were performed in the 2.2--10~keV energy band, to avoid large gradients of the instrument
transfer function due to the instrumental edges (1.8--2.2~keV) and to the soft X-ray photoelectric cut-off.
Below the energy of the latter the X-ray spectrum is dominated by reprocessing and scattering of the nuclear
radiation by photoionised gas on scales as large as a few hundreds parsecs \cite{bianchi03}. High-resolution
spectroscopy of this component will be discussed elsewhere (Labiano et al., in preparation).

In this paper: energies are quoted in the source rest frame; errors are quoted at the 90\% confidence level
for one interesting parameter;
the following cosmological parameters have been adopted in the calculation of the luminosities:
$H_0$=70~km~s$^{-1}$~Mpc$^{-1}$, $\Lambda_0$=0.73,
$\Omega_m$=0.27 \cite{bennett03}.

\section{Spectral analysis}

\subsection{Phenomenological fits on individual observations}

As first step, we fit the pn spectrum of each individual XMM-Newton observation with the same
phenomenological model. Following the approach as in Matt et al. (2001),
the model is constituted by a photoelectrically absorbed power-law, a photoelectric absorption edge and three
narrow iron emission lines, modelling the K$_{\alpha}$ fluorescence of Fe{\sc i}
and of recombination lines of
Fe{\sc xxv} and Fe{\sc xxvi}. The Fe{\sc i} and Fe{\sc xxvi} are actually each close doublets, which
cannot be resolved by the EPIC cameras. Their centroid energy has been fixed
to the weighed mean of the corresponding doublet components: 6.399~keV and 6.966~keV, respectively.
The structure of the Fe{\sc xxv} line is instead more complex, with the resonance (6.7002~keV), two intercombination
(6.6821~keV, and 6.6673~keV) and the forbidden components (6.6364~keV) all potentially contributing.
In the fit of each observation we left originally the Fe{\sc xxv}
centroid energy free to vary. We obtained lower limits
consistent with the energy of the resonance line (and inconsistent with the other components)
in all observations, except in XMM8, where
$E^{FeXXV}_c = 6.654\pm^{0.020}_{0.017}$~keV [$I = (3.7 \pm 1.0) \times 10^{-5}$~photons~cm$^{-2}$~s$^{-1}$].
In all the subsequent models we fixed the centroid energy of the
Fe{\sc xxv} line to the value of the resonance component, except for XMM8 for which we used the value of the
forbidden component.
Finally, the energy of the edge was constrained to be $\approxgt$7.11~keV, corresponding to
photoionisation of iron.
 
The Fe{\sc i} line 1-$\sigma$
width (only ``width'' hereafter) is $46 \pm^{11}_{14}$~eV. This measurement is consistent with the
width of the Manganese K$_{\alpha}$ and K$_{\beta}$ pn calibration emission lines measured in observations
close to the 2009 NGC~5506 observations: $\sigma_{cal} = 42 \pm 8$~eV, and $\sigma_{cal} = 47 \pm 8$~eV
in Obs.\#0410780601 (August 9, 2008) and  Obs.\#0412580401 (January 4, 2009),
respectively. The measurement is thus consistent with the instrumental resolution.
We have consequently used this value for all the
nominally unresolved astrophysical emission lines in the models discussed in this paper.

\begin{figure}
  \includegraphics[height=85mm,angle=-90]{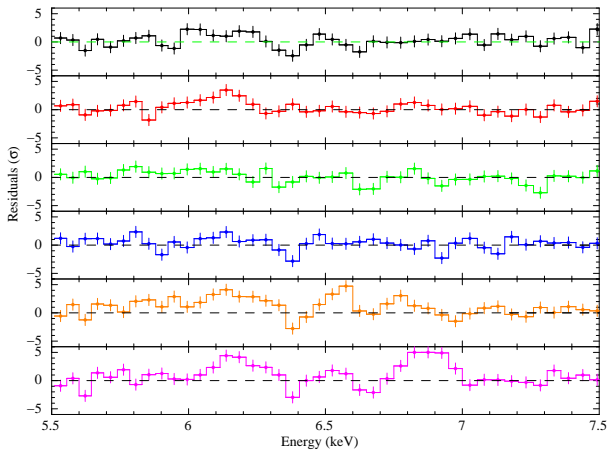}
  \caption{5.5-7.5~keV residuals in units of standard deviations when the best-fit phenomenological model
is applied to the
2.2--12~keV EPIC-pn spectra of the 2004-2009
XMM-Newton observations of NGC~5506.}
  \label{fig1}
\end{figure}

In Fig.~\ref{fig1} we show the residuals against the best-fit phenomenological
model
in the 5.5--7.5~keV energy band (the residuals are flat outside this interval). In the longest
observations significant
wave-like residuals are present around the nominal energy of the Fe{\sc i} fluorescent K$_{\alpha}$ line.
Hints of features with a similar shape are visible in other observations with a lower statistics.
These features can be 
interpreted as being due to a broad ($\sigma \simeq$300~eV) excess emission feature, which the narrow
Fe{\sc i} Gaussian
profile is unable to account for. 

\begin{figure*}
\hbox{
  \includegraphics[height=85mm,angle=90]{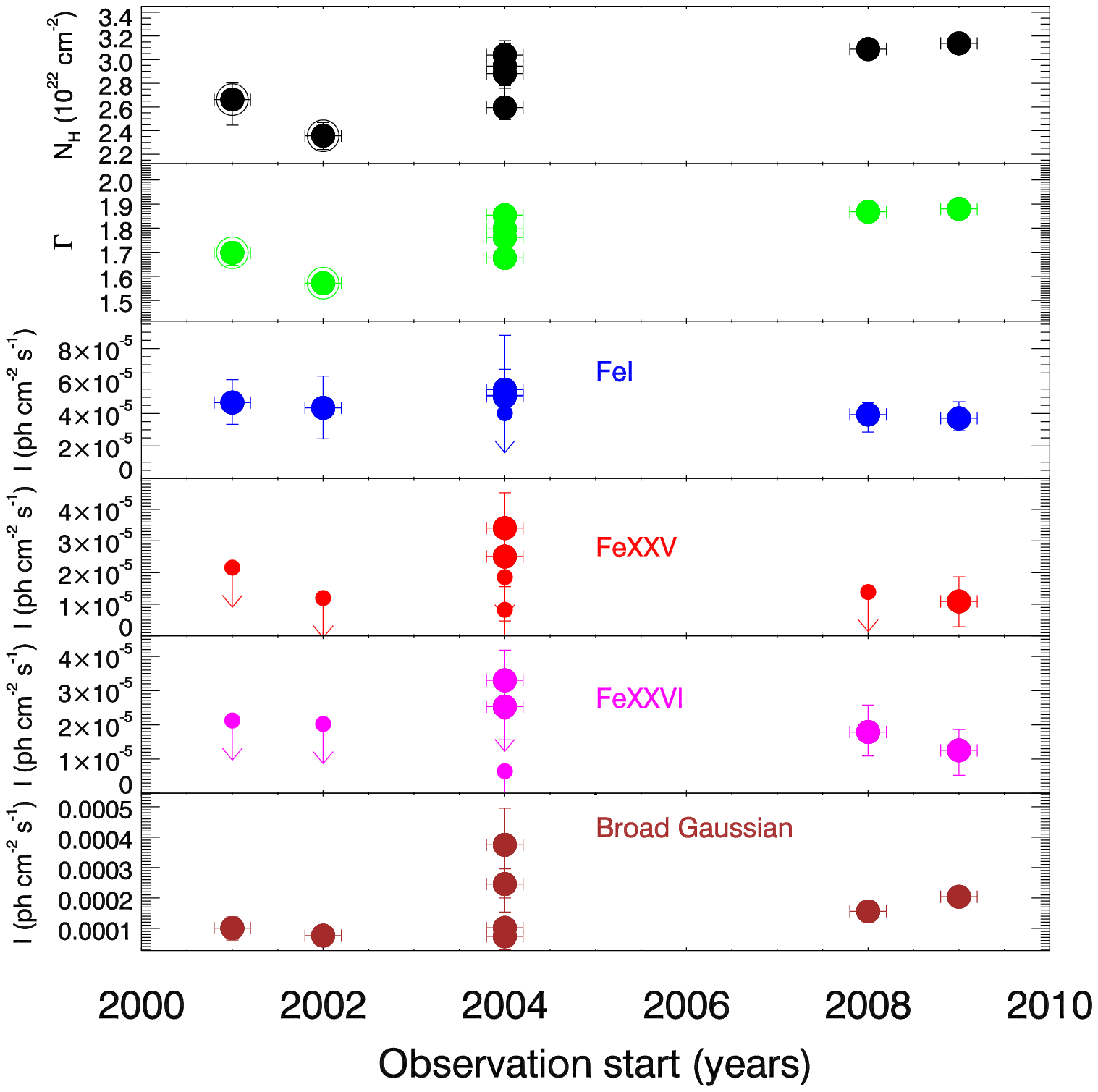}
  \hspace{0.5cm}
  \includegraphics[height=85mm,angle=90]{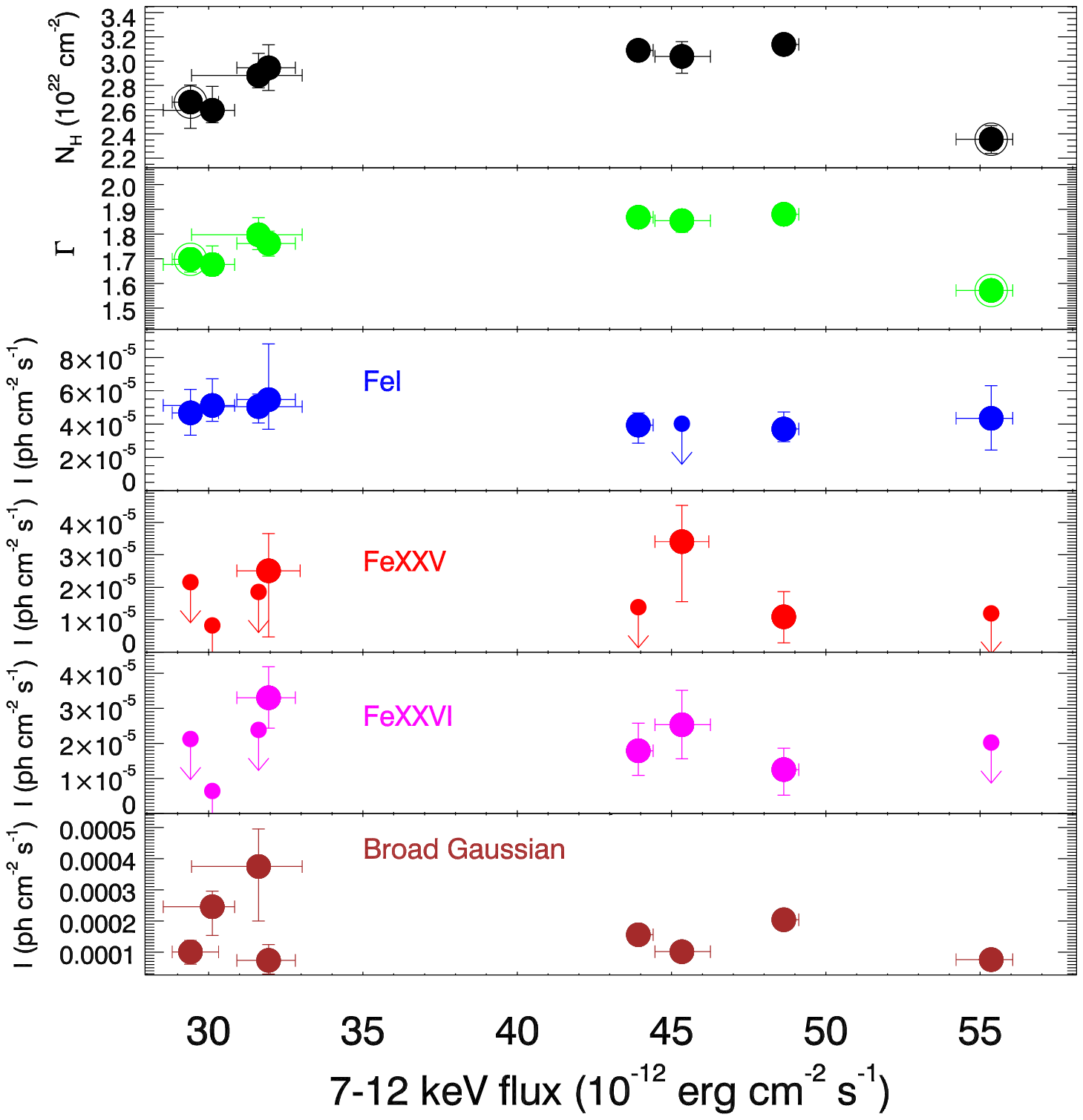}
}
  \caption{Best-fit parameters of the phenomenological fits of individual NGC~5506 XMM-Newton
  observations as a function
  of time ({\it left panel}) and of the 7--12~keV absorption-corrected flux ({\it right panel}):
  from {\it top} to {\it bottom}: photon index, column density, intensity of the Fe{\sc i}, Fe{\sc xxv},
  Fe{\sc xxvi} narrow emission lines and of the broad Gaussian. Continuum measurement data points corresponding
  to pn observations in Large Window mode are surrounded by a {\it large circle}.}
  \label{fig4}
\end{figure*}

In Fig.~\ref{fig1} we do not show the residuals for the first two observations (XMM1 and XMM2).
The event pattern distribution measured by the SAS task {\tt epatplot} suggests that they
might be marginally affected
by pile-up due to the different instrument mode (Large Window instead of Small Window).
This may also explain why XMM2 - which corresponds to the highest X-ray flux measured during
our campaign - exhibits the flattest X-ray spectral index. This is illustrated in
Fig.~\ref{fig4}, where the best-fit parameters of the phenomenological model in each
individual fit as a function of time and of the 7--12~keV absorption-corrected flux are shown, once
the excess emission around 6~keV is fit with a further Gaussian profile whose width was left
free to vary.

The intensity of
the narrow Fe{\sc i} line is constant. Its mean normalised dynamical range is $\le 27\%$.
Evidence for variability of the recombination He- and H-like line intensity
is also marginal. Beside the possible change of ionisation state of the Fe{\sc xxv} in XMM8,
it is limited to a nominal difference by a factor of at least five between
two measurements of the Fe{\sc xxvi} line during the 2004 monitoring campaign. Given
the uncertainties in the spectral deconvolution at $\simeq$7~keV, we consider this
evidence for variability as only tentative, to be confirmed by high-resolution
measurements.

We have applied the same phenomenological model also to the combined spectra extracted
from the MOS cameras. The parameters of the continuum are different. The ratio between the 2--10~keV
MOS and pn fluxes (weighted mean, $\langle F_{MOS}/F_{pn} \rangle$=5.5\%; $\sigma_{F_{MOS}/F_{pn}}$=2.7\%)
is consistent with known systematic calibration uncertainties in the absolute flux determination
\cite{stuhlinger08}. However, the average difference between the spectral index
($\langle \Delta \Gamma \rangle$=0.23; $\sigma_{\Delta \Gamma}$=0.07) and the column density
($\langle \Delta N_H \rangle$=4.9$\times 10^{21}$~cm$^{-2}$; $\sigma_{\Delta N_H}$=2.0$\times 10^{21}$~cm$^{-2}$)
is larger then expected. The difference is larger for higher fluxes. The
continuum spectral parameters measured by the EPIC cameras are instead well consistent for observations
whose 2--10~keV flux is lower than $\simeq$8$\times 10^{-11}$~erg~cm$^{-2}$~s$^{-1}$.
Visual inspection of the MOS pattern fraction distribution unveiled an excess (deficit) of
double (single) events by 15--30\% above 4~keV in higher flux observations. The above pieces of evidence
indicate that many of the NGC~5506 MOS spectra could be still affected by pile-up.
Consequently, we will not discuss the MOS spectra in the rest of the paper. However, we will show in Sect.~3.2 that
the main result of this paper - the ultimate discovery of a broad component of the iron
K$_{\alpha}$ line - is measured in {\it both} the pn and the MOS cameras with comparable properties.

\subsection{Physically-motivated combined fits of all observation spectra}

\subsubsection{Description of the models}

Motivated by the results presented in the previous Section, we
enriched the phenomenological model by adding physically-motivated components to account
for features detected in the spectral residuals. We have investigated
three main scenarios:

\begin{itemize}

\item Scenario \#1: relativistically broadened K$_{\alpha}$ iron line and non-relativistic
Compton reflection. The former component has been modelled through the code {\tt kyrline}
\cite{dovciak04}, which calculates the profile of an emission line produced in an
axisymmetric accretion disk around
a black hole. The Compton-reflection component (model {\tt pexrav}; Magdziarz \& Zdziarski 1995) is
primarily associated in this
scenario to the narrow core of the iron K$_{\alpha}$ line, although it may include a contribution by the
accretion disk as well, which the moderate energy resolution of the EPIC cameras is unable to
disentangle.

\item Scenario \#2: non-relativistically broadened K$_{\alpha}$ iron line and Compton reflection.
The former component has been modelled with a simple Gaussian profile

\item Scenario \#3: relativistically broadened emission line and continuum reflection dominated by
a relativistically blurred component.
The latter component has been modelled by convolving {\tt pexrav} with the relativistic kernel
{\tt kyconv} \cite{dovciak04} and assuming a constant ratio between the normalisation of the
primary continuum and that of the Compton reflection ($R$, see Tab.~\ref{tab2}).

\end{itemize}

{\tt kyrline} belongs to a suite of models based on a common ray-tracing subroutine aiming at
describing the X-ray emission of black-hole accretion disks in the strong gravity regime
\cite{dovciak04}. The line profile depends on the following parameters:
a) the black hole spin $a$ (comprised between 0 and 0.9982 in dimensionless relativistic units);
b) the inner ($r_{in}$) and outer ($r_{out}$) radius of an annular region on the disk where
the line photons are emitted; c) the index $\alpha$ of the
radial dependence of the emissivity per unit area, $\kappa$, in the local frame
comoving with the disk: $\kappa (r) \propto r^{-\alpha}$; d) the
``disk inclination'' $\imath$, i.e. the angle between the normal to the disk plane and
the line-of-sight; e) the rest frame
energy (assumed monochromatic) of the photons emitted by the disk, $E_c$.
Although the line normalisation in {\tt kyrline} is expressed in units of
photons~cm$^{-2}$~s$^{-1}$ integrated over the whole profile, we will also use the line
Equivalent Width (EW), i.e. the line intensity normalised to the underlying continuum at
6.4~keV, to ease comparison against theoretical predictions \cite{matt92}. The convolution
function {\tt kyconv} uses
the same relativistic kernel as {\tt kyrline}.

The continuum reflection components replace the photo-absorption edge in the phenomenological model.
We always linked together the spectral index
of the Compton-reflected continuum and that of the primary emission.
Moreover, we have added two narrow-band spectral components to the non-relativistic emission line complex, modelling the:
a) iron K$_{\alpha}$ fluorescent iron line Compton-shoulder \cite{sunyaev96,matt02} through a
rectangular box function
comprised between 6.24 and 6.4~keV, and whose intensity was constrained to be not larger then 20\% the
intensity of the corresponding narrow K$_{\alpha}$ line (the fits determine
only an upper limit on this parameter as large as the maximum allowed value);
b) the K$_{\beta}$ line associated with the iron K$_{\alpha}$ narrow component,
modelled with a Gaussian profile with energy fixed at 7.058~keV and intensity constrained not to exceed
16\% of the K$_{\alpha}$ intensity \cite{molendi03}.

In principle, only model parameters describing physical quantities not expected to change among the observations
(or demonstrated to remain constant in the phenomenological analysis)
have been constrained to have the same value for all fitted spectra. They are:
the intensities of the narrow Fe{\sc i}, Fe{\sc xxv} and Fe{\sc xxvi} lines; the metallicity (assumed solar;
Anders \& Grevesse 1989); the
high-energy cut-off of the intrinsic primary continuum (held fixed to 130~keV in accordance to the
results of the BeppoSAX observation; Bianchi et al. 2003);
the inclination of the disk and of the distant non-relativistic reflector
(the latter held fixed to 45$^{\circ}$); and the black hole spin. In Scenario\#1 and \#3 varying the black hole spin from
Schwarzschild to maximally rotating yields a variation of the $\chi^2$ lower than 0.5. We have therefore
fixed this parameter to the value corresponding to the latter solution ($a = 0.998$),
favoured at face value. Once these assumptions
were made, a first series of fits
showed that many of the remaining free parameters
were consistent with being constant across all the fitted observations. We have
therefore forced them to assume the same value in all the fits. In Scenarios~\#1 and \#3 they are: a) the
inner and outer radius of the X-ray emitting disk. Initially we have
fixed the latter to 400$r_g$, because this parameter is generally unconstrained
when the line profile is strongly relativistic. We will review this assumption in Sect.~3.2.2.
The former has been constrained
throughout the paper to coincide with the innermost
stable circular orbit of an accretion disk co-rotating with a black hole of spin $a$; b)
the index of the power-law disk emissivity
radial dependence, $\alpha$; c) the energy of the relativistically broadened line in the source rest frame, $E_c$.
In Scenario~\#2 they are all the parameters describing the broad
Gaussian profile.

\subsubsection{Results}

\begin{figure}
  \includegraphics[height=80mm,angle=-90]{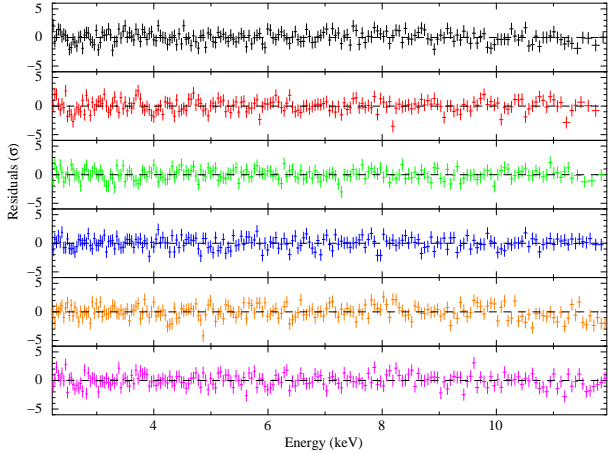}
  \caption{Residuals in units of standard deviation when the Scenario\#3 best-fit model is
applied to the 2004-2009 EPIC-pn spectra. From {\it top} to {\it bottom}: XMM3, XMM4, XMM5, XMM6, XMM7, XMM8.
Y-axis range and colours are the same as in Fig.~\ref{fig1}}
  \label{fig2}
\end{figure}

\begin{figure*}
\hbox{
  \includegraphics[height=80mm,angle=-90]{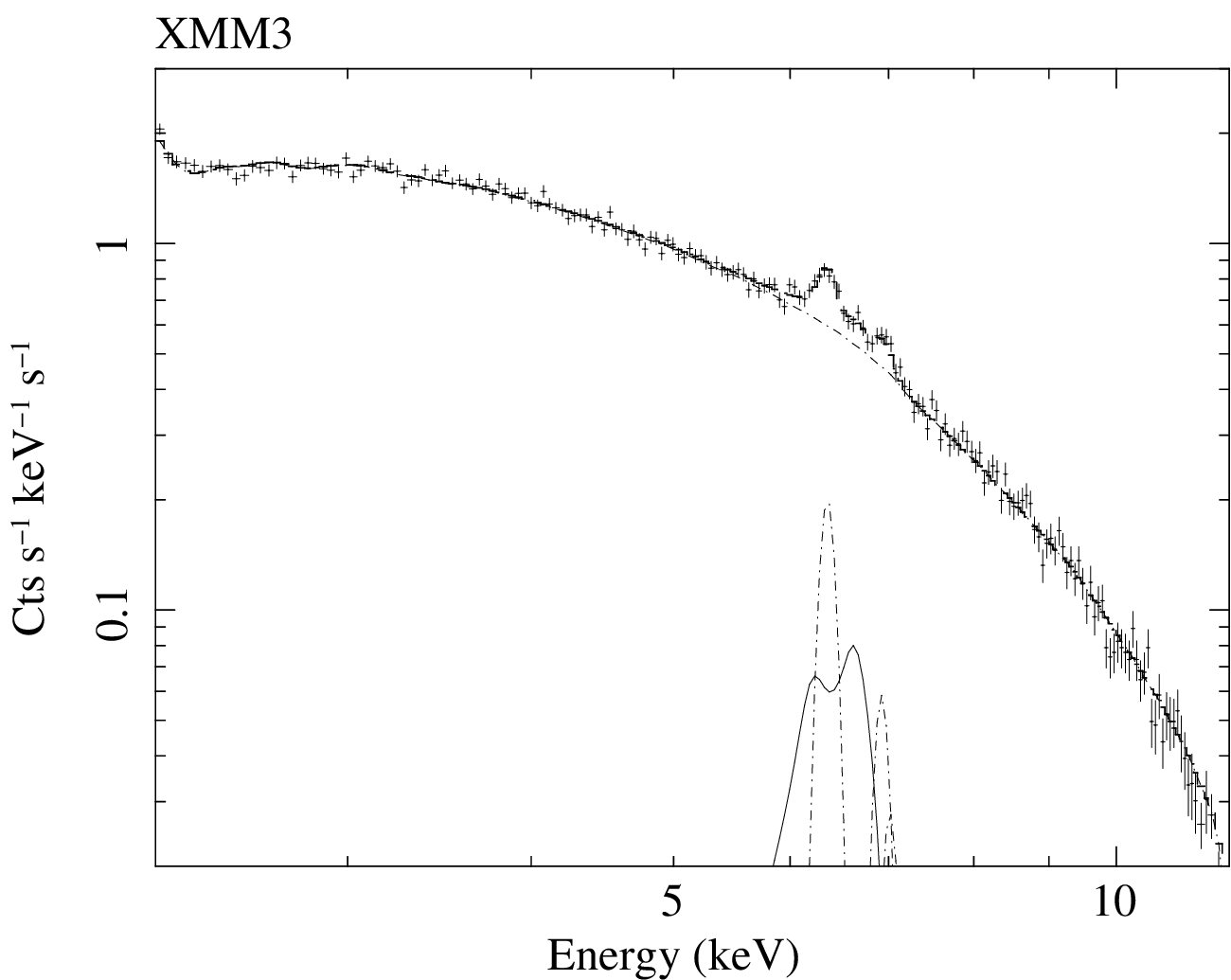}
  \includegraphics[height=80mm,angle=-90]{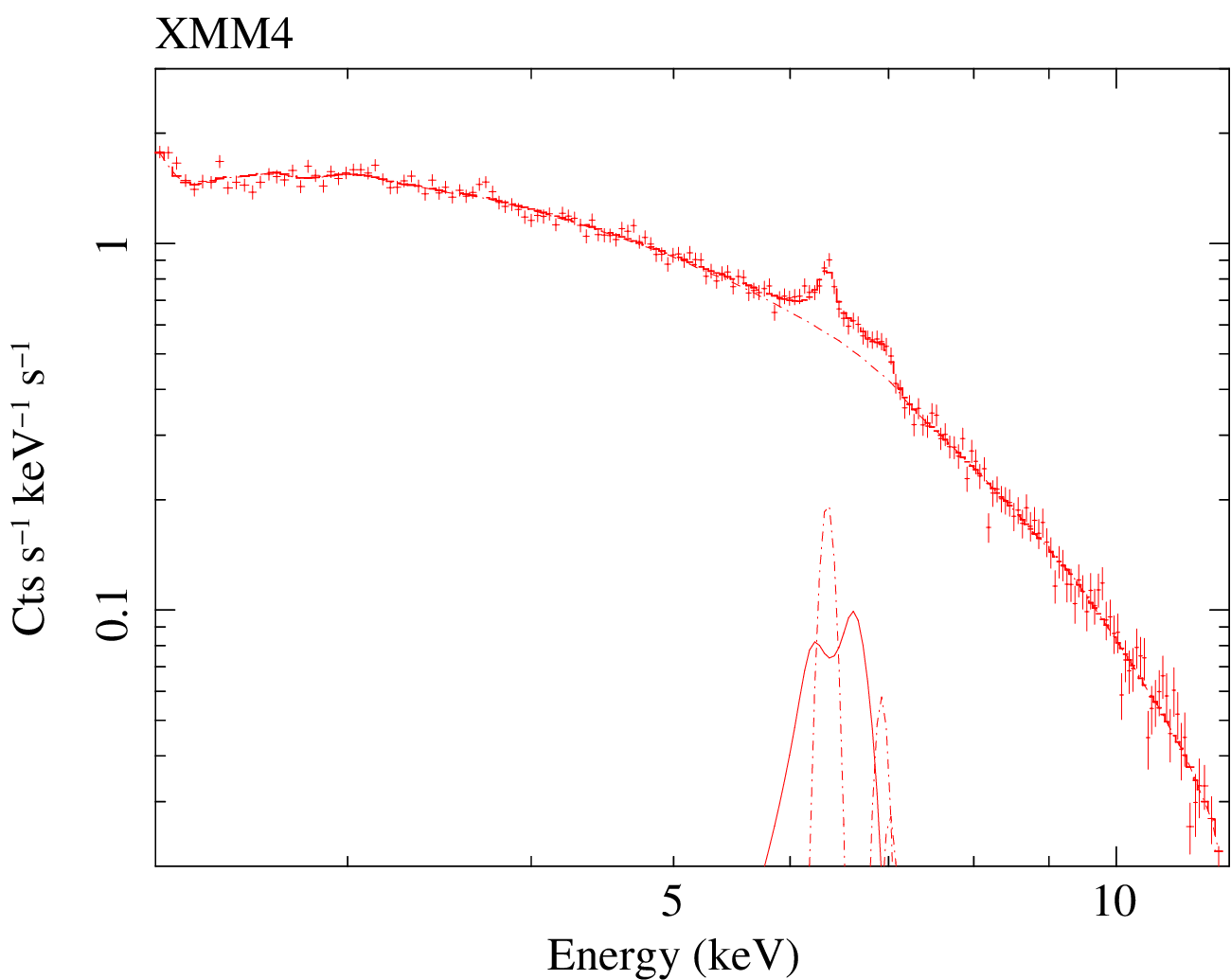}
}
\hbox{
  \includegraphics[height=80mm,angle=-90]{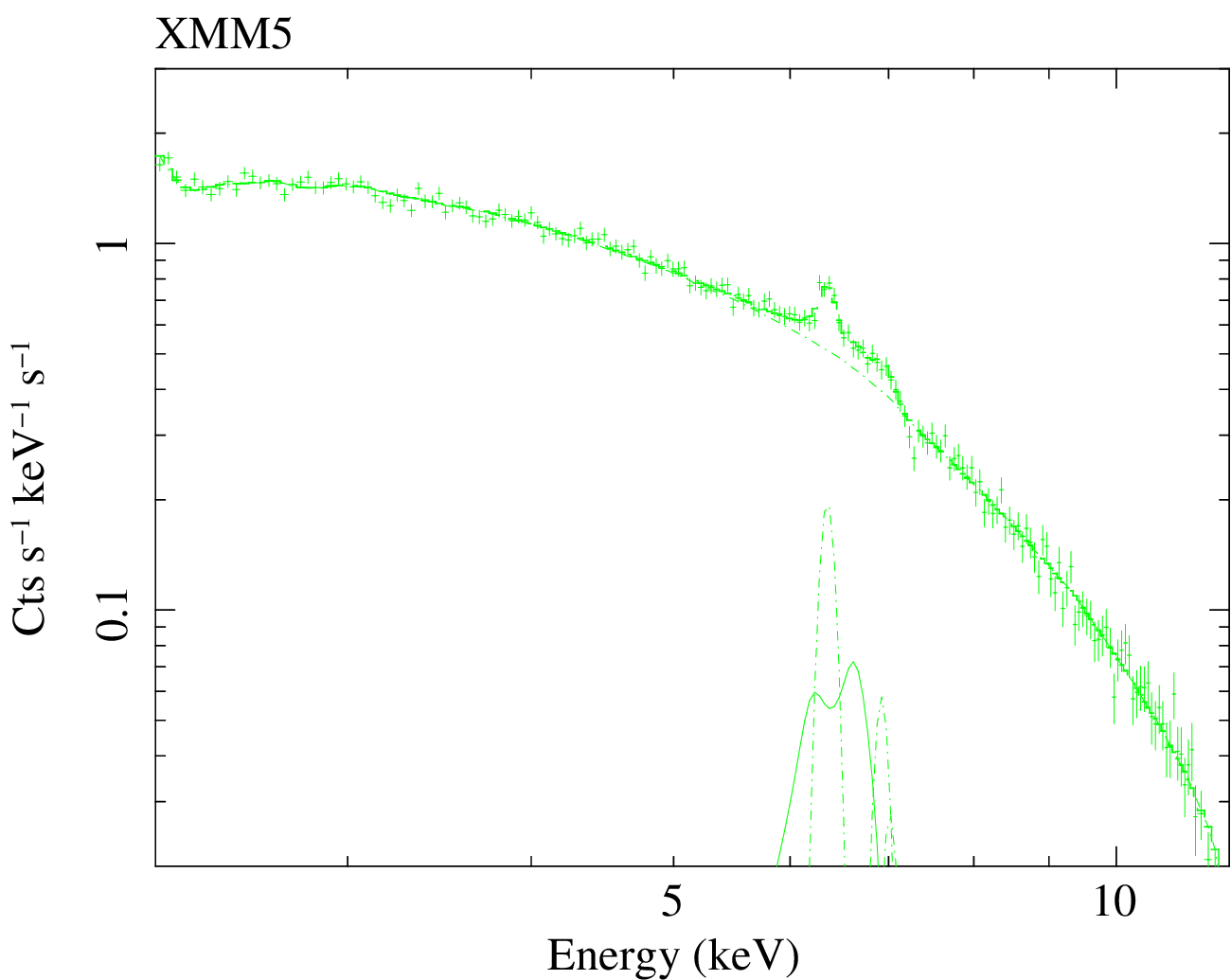}
  \includegraphics[height=80mm,angle=-90]{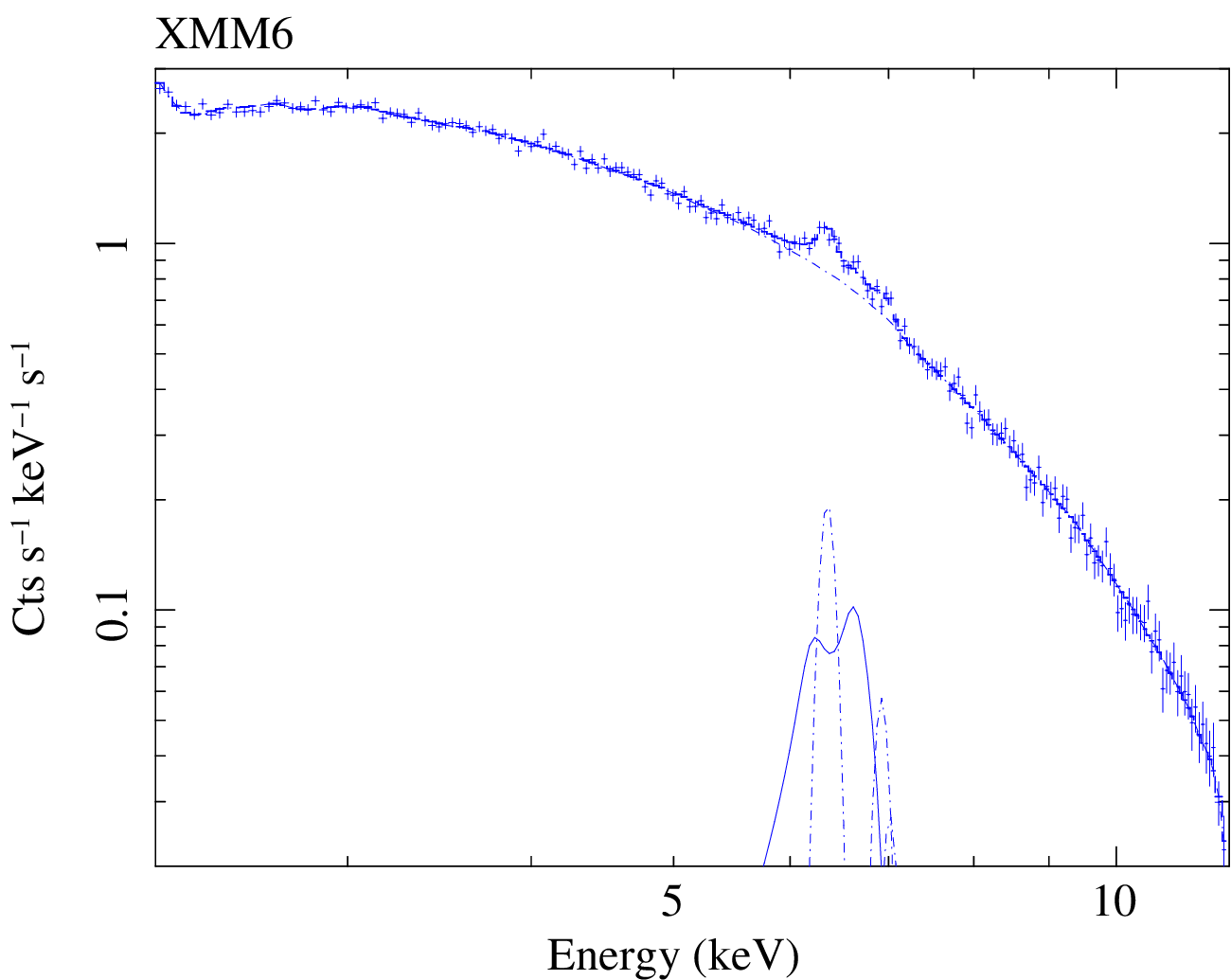}
}
\hbox{
  \includegraphics[height=80mm,angle=-90]{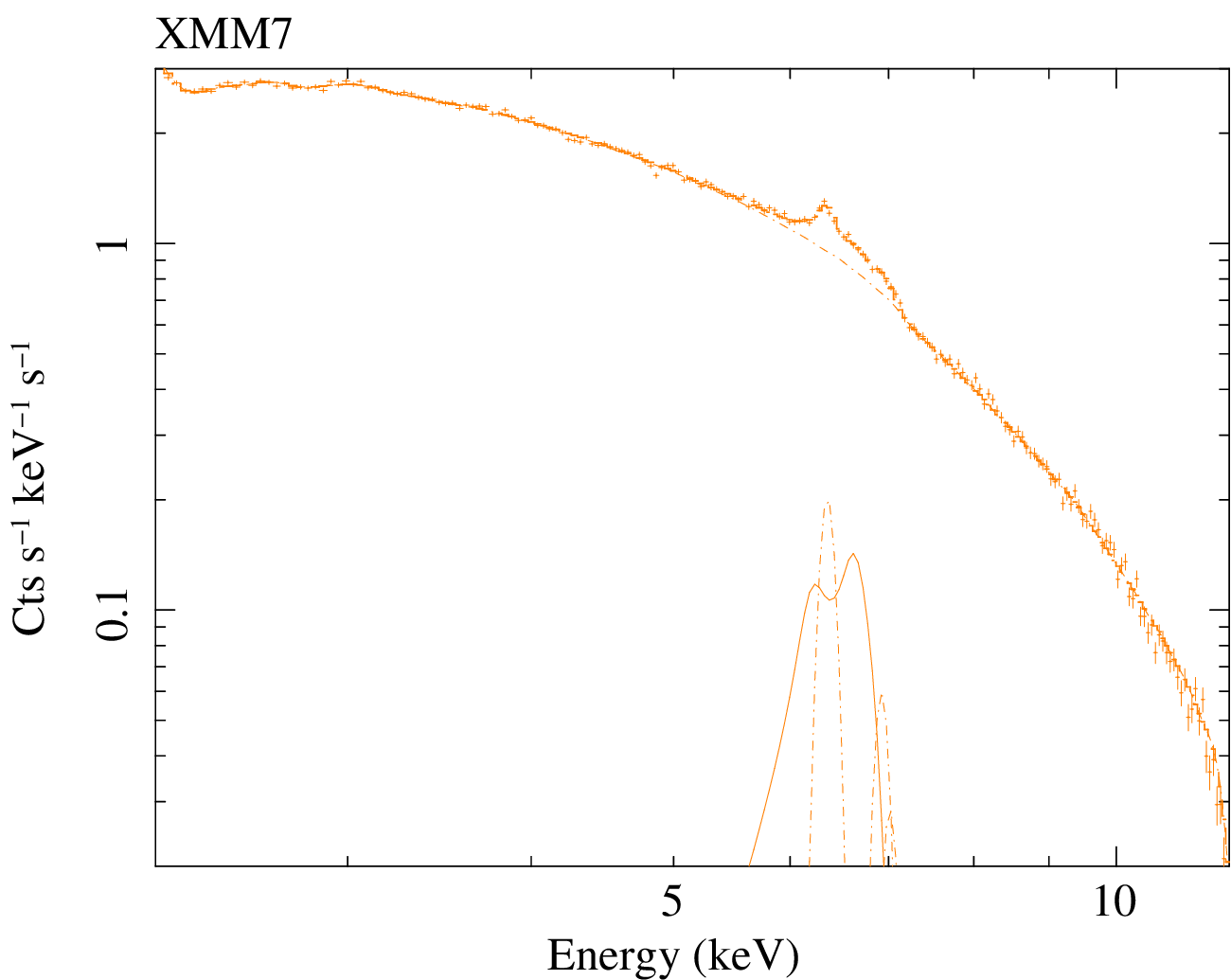}
  \includegraphics[height=80mm,angle=-90]{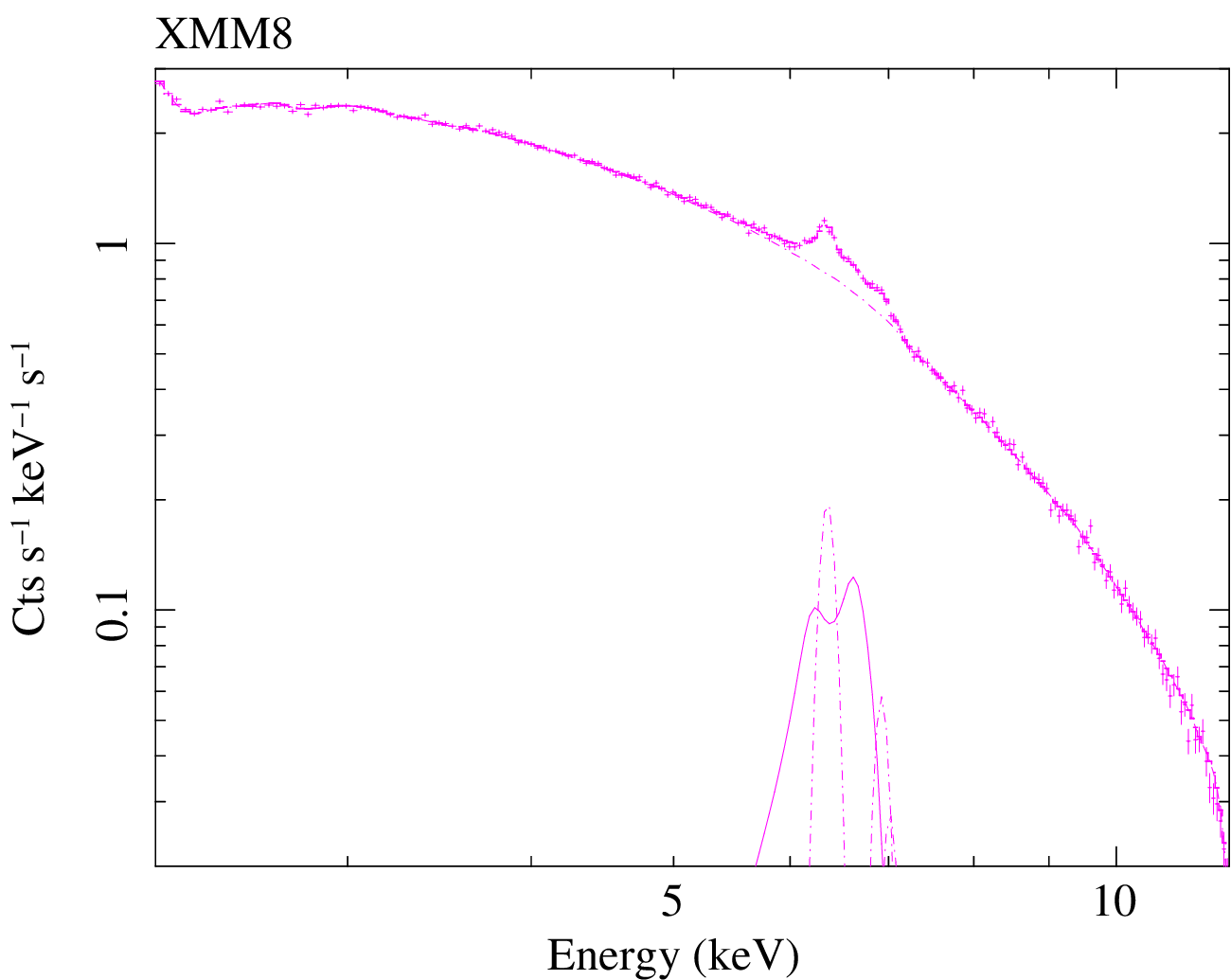}
}
  \caption{EPIC-pn NGC~5506 spectra in observations XMM3 to XMM8 ({\it crosses}). The {\it lines} indicate
different model components: {\it dot-dashed}: continuum; {\it long dot-dashed}: narrow lines;
{\it solid}: broad line.}
  \label{fig9}
\end{figure*}

The Scenarios described above yield comparable fit qualities: $\chi^2/\nu$=1265.3/1161,
1280.9/1161, and 1258.1/1161 for Scenario~\#1, \#2, and \#3, respectively. 
These $\chi^2$ are significantly better than those yielded by the same models {\it without} the
broad line: 1479.5/1169, 1516.2/1169, and 1581.6/1166, respectively.
No further structure is visible in the residuals (Fig.~\ref{fig2}).
We therefore concluded that no additional astrophysical
component is required.
In Tab.~\ref{tab2} we summarise the best-fit results for the non-relativistic components in Scenario\#3.
\begin{table}
  \caption{Best-fit results for the parameters of the non-relativistic continuum components in Scenario~\#3
    (except parameter $EW_{FeI}$ which refers to Scenario\#1).}
  \label{tab2}
  \begin{tabular}{lccc}
  \hline
\multicolumn{4}{l}{Spectrum dependent parameters} \\
Obs.\# & $N_H$ & $\Gamma$ & 2--10~keV Flux$^a$ \\
& ($10^{22}$~cm$^{-2}$) & & ($10^{-12}$~erg~s$^{-1}$~cm$^{-2}$) \\
XMM3       & $2.86 \pm^{0.11}_{0.14}$ & $1.778\pm^{0.014}_{0.025}$ & $82.5  \pm^{0.6}_{1.2}$ \\
XMM4       & $2.94 \pm^{0.12}_{0.14}$ & $1.777\pm^{0.013}_{0.033}$ & $79.4  \pm^{1.2}_{1.1}$ \\
XMM5       & $2.65 \pm^{0.12}_{0.14}$ & $1.771\pm^{0.014}_{0.027}$ & $71.4  \pm^{0.7}_{1.3}$ \\
XMM6       & $2.99 \pm^{0.08}_{0.12}$ & $1.844\pm^{0.016}_{0.015}$ & $120.0 \pm^{0.6}_{1.3}$ \\
XMM7       & $3.08 \pm^{0.04}_{0.03}$ & $1.864\pm 0.009$           & $132.5 \pm^{1.3}_{0.9}$ \\
XMM8       & $3.04 \pm^{0.06}_{0.05}$ & $1.862\pm^{0.010}_{0.006}$ & $118.7 \pm^{0.3}_{0.7}$ \\
\hline
\multicolumn{4}{l}{Spectrum independent parameters} \\
\multicolumn{4}{l}{$I_{FeI}^b = (5.9 \pm 0.6) \times 10^{-5}$~erg~cm$^{-2}$~s$^{-1}$} \\
\multicolumn{4}{l}{$I_{FeXXVr}^b \le 1.0 \times 10^{-5}$~erg~cm$^{-2}$~s$^{-1}$} \\
\multicolumn{4}{l}{$I_{FeXXVf}^b \le 1.4 \times 10^{-5}$~erg~cm$^{-2}$~s$^{-1}$} \\
\multicolumn{4}{l}{$I_{FeXXVI}^b = (2.0 \pm 0.7) \times 10^{-5}$~erg~cm$^{-2}$~s$^{-1}$} \\
\multicolumn{4}{l}{$R \le 0.1$$^c$} \\
\multicolumn{4}{l}{$EW_{FeI} \ge 1.0$~keV$^d$} \\
 \hline
\end{tabular}

\noindent
$^a$corrected for absorption. XMM1: $70.8 \pm 1.0$; XMM2: $121.1\pm^{1.4}_{1.1}$ in the same units;
$^b$intensity of the narrow emission lines;
$^c$ratio between the primary continuum and the relativistically blurred reflection normalisation;
$^d$ Equivalent Width of the narrow component of the Fe{\sc i}
fluorescent line against its own reflection continuum.
\end{table}
The values for the other scenarios are consistent with them within the statistical uncertainties.
In Fig.~\ref{fig9} we show the spectra of observations XMM3 to XMM8 together with the components
of their best-fit models.

\begin{figure}
  \includegraphics[height=70mm,width=80mm]{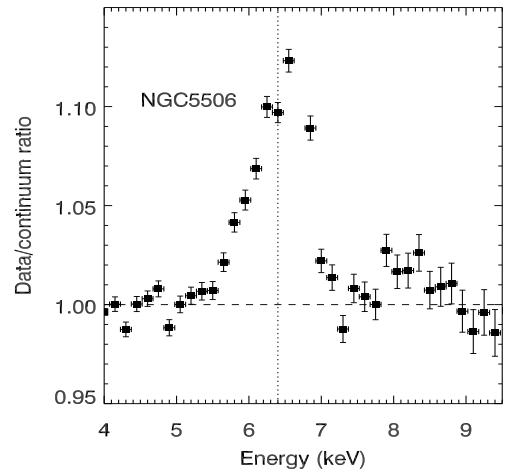}
  \caption{Stacked profile of the
  relativistically broadened iron line. The {\it dashed line} indicates
  the observed frame energy of the neutral iron K$_{\alpha}$ fluorescent line.}
  \label{fig6}
\end{figure}

The properties of the relativistic line can be summarised as follows (see Tab.~\ref{tab5}):
\begin{table}
\caption{Best-fit values of the K$_{\alpha}$ iron relativistically broadened profile in Scenarios~\#1 and
\#3 for $r_{out}$=400$r_g$. The {\it rightmost column} contains the statistical uncertainties to
be added if $r_{out}$ is allowed to vary.}
\label{tab5}
\begin{tabular}{lccc} \hline
& Scenario~\#1 & Scenario~\#3 & Additional error \\
&  &  & if $r_{out}$ is left free \\
$E_c$ (keV) & $6.48 \pm 0.02$ & $6.48 \pm 0.02$ & $\pm^{0.02}_{0.00}$ \\
$\alpha$ & $1.9 \pm 0.3$ & $1.9 \pm 0.3$ & $\pm^{0.2}_{0.1}$ \\
$\imath$$^{\circ}$ & $42 \pm^2_6$ & $37\pm^6_5$ & $\pm^{11}_{4}$ \\
$r_{in}$ ($r_g$) & $15 \pm^9_{12}$ & $15 \pm^{11}_9$ & $\pm^{7}_{1}$ \\
\hline
\end{tabular}
\end{table}
a) the rest frame centroid energy is consistent with neutral or
moderately ionised iron
($E_c \simeq 6.48$~keV). Fixing the source frame centroid energy to the value expected for
neutral iron in, {\it e.g.}, Scenario\#3 yields $\chi^2=$1276.0/1160~dof;
b) the disk is slightly more inclined ($\imath \simeq 40^{\circ}$)
than typically measured in unobscured Seyfert galaxies \cite{nandra97,nandra07}, and;
c) the line profile is only mildly relativistic
as parametrised by the moderately flat ($\alpha \simeq$1.9)
radial emissivity profile. The line emitting region extends
close to the innermost stable orbit. The innermost radius is constrained
between 3 and 25~$r_g$.
In Fig.~\ref{fig6} we show the profile of the relativistically
broadened iron line, obtained by stacking together the source frame residuals of each individual observation against
the best-fit model in Scenario\#3 once the relativistically broadened emission line had is removed from the
model. The profile is not significantly skewed. This is consistent with the similar fit quality that
relativistic and non-relativistic scenarios yield.
A small ``dip'' in the profile coincide exactly with the rest frame energy of the
K$_{\alpha}$ neutral iron fluorescent line. It could be due to an incorrect disentangling 
between the narrow and the broad profile, whereby the fit preferentially attributes photons
in this energy range to the former. We estimate that the intensity of the narrow component should
be lowered by at least 50\% for the dip to disappear.
The effect of this potential systematic
uncertainty on the determination of the relativistic line profiles $\alpha$ and
$r_{in}$ is negligible, because they are mostly sensitive to the red wing. On the other hand, centroid
energy and inclination angle could be affected.
Assuming $I_{FeI} = 3 \times 10^{-5}$~photons~cm$^{-2}$~s$^{-1}$, they change by $\Delta E_c \simeq 30$~eV
and $\Delta \imath \simeq 4^{\circ}$, respectively. Nonetheless, the $\chi^2$ significantly worsens
(1361.9/1160~dof in Scenario\#3). This hypothesis is then inconsistent with the data.

We come back now to the original restriction imposed to the value of $r_{out}$. The flat value of
$\alpha$ implies that a not negligible contribution to the broad line profile comes also from
large disk radii. We have therefore repeated the fits in Scenario\#3 after allowing $r_{out}$ to be a free
parameter. At the 90\% confidence level for one interesting parameter $r_{out}$ is constrained to
be comprised between 250 and 900$r_g$. The best-fit value is still around 400$r_g$, hence the
best-fit values do not change. However the broad line
parameters errors are augmented by a small amount (Tab.~\ref{tab5}).

In Scenario~\#2, the centroid energy of the Gaussian profile
is consistent with fluorescence from neutral or moderately ionised iron:
$E_c = 6.45 \pm^{0.04}_{0.05}$~keV. The line width is $\sigma_b = 330 \pm 40$~eV.

\begin{figure}
  \includegraphics[height=80mm,angle=-90]{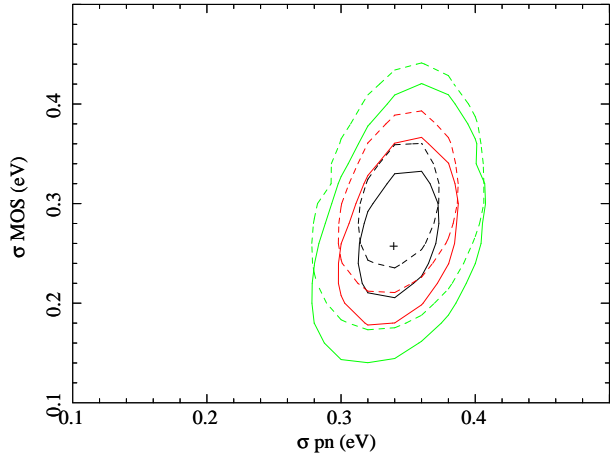}
  \caption{Iso-$\chi^2$ contour plots for the intrinsic width of a Gaussian profile fitting
    the broad component of the iron K$_{\alpha}$ line in the NGC~5506 merged spectra. The contours correspond to
    the 68\%, 90\% and 99\% confidence level for two interesting parameters, respectively.
    {\it Solid lines}: pn versus MOS1; {\it dashed lines}: pn versus MOS2.}
  \label{fig5}
\end{figure}

We have checked the robustness of the broad iron line profile against possible data reduction and/or
undiagnosed calibration problems.
A fit with Scenario\#2 on a merged pn spectrum with single and double events yields the following best-fit
broad Gaussian profile parameters: $E_c = 6.47 \pm^{0.04}_{0.03}$~keV, $\sigma_b = 360 \pm^{40}_{30}$~eV,
and $I_b = (1.47 \pm 0.14) \times 10^{-4}$~ph~s$^{-1}$~cm$^{-2}$. They agree within the statistical
uncertainties with the best-fit parameters obtained on a merged pn spectrum extracted with single events only:
$E_c = 6.41 \pm 0.04$~keV, $\sigma_b = 360 \pm^{50}_{70}$~eV,
and $I_b = (1.55 \pm 0.18) \times 10^{-4}$~ph~s$^{-1}$~cm$^{-2}$.
More importantly, the width measured by the pn spectrum agrees
with that simultaneously measured by the MOS cameras (Fig.~\ref{fig5}).

\subsubsection{Variability}

The EW of the broad component of the iron K$_{\alpha}$ line is consistent with being constant
among the XMM-Newton observations discussed in this paper (Tab.~\ref{tab4}). On the other hand, its
\begin{table}
 \centering
  \caption{Measurements of the broad iron K$_{\alpha}$ line EW (in eV).
           }
  \label{tab4}
  \begin{tabular}{lccc}
  \hline
Obs.\# &  Scenario~\#1 & Scenario~\#2 & Scenario~\#3 \\
  \hline
XMM3       & $130 \pm^{40}_{20}$ & $130 \pm 30$           & $120 \pm^{30}_{20}$ \\
XMM4       & $180 \pm^{30}_{40}$ & $160 \pm 30       $    & $150 \pm^{30}_{20}$ \\
XMM5       & $140 \pm  30      $ & $130 \pm 30$           & $120 \pm 30       $ \\
XMM6       & $122 \pm^{17}_{26}$ & $100 \pm^{30}_{20}$    & $110 \pm 20       $ \\
XMM7       & $152 \pm^{11}_{12}$ & $137 \pm^{13}_{15}$    & $132 \pm^{13}_{17}$ \\
XMM8       & $149 \pm^{18}_{15}$ & $143 \pm^{12}_{16}$    & $131 \pm 15       $ \\
\hline
XMM78$_1$ & $180 \pm^{20}_{30}$ & $140 \pm^{30}_{20}$  & $160 \pm 20       $ \\
XMM78$_2$ & $117 \pm^{13}_{26}$ & $90  \pm^{30}_{20}$  & $100 \pm^{30}_{20}$ \\
XMM78$_3$ & $110 \pm^{30}_{20}$ & $90  \pm 20       $  & $101 \pm^{18}_{20}$ \\
XMM78$_4$ & $120 \pm 20       $ & $100 \pm 20$         & $109 \pm^{16}_{19}$ \\
 \hline
\end{tabular} 
\end{table} 
{\it intensity} is not constant.
A fit on the broad line
intensity versus time relation with a constant yields a $\chi^2/\nu$=11.4/5 (versus 2.1/5 for the EW).
This indicates that in the first approximation
the broad line intensity follows the variation of the underlying continuum.

The absorption-corrected 2--10~keV flux measured during the XMM-Newton observations covers a
dynamical range $\simeq$1.9. This is less than a factor of two smaller than measured during the 6-years long
intense X-ray monitoring with the RXTE/PCA in the same energy band ($\simeq$3.5; Uttley \& McHardy 2005).
The longest time difference between two observations, whose continuum flux differs by more then
27\% (the upper limit on the percentage fractional variability of the intensity of the
iron K$_{\alpha}$ fluorescent line narrow component) is $\simeq$7.92~years (XMM8 versus XMM1).

The photon index in observations XMM6 to XMM8 (2--10~keV absorption corrected flux,
$F_{2-10} \ge 1.19 \times 10^{-10}$~erg~s$^{-1}$~cm$^{-2}$) is systematically steeper than in the others
(by $\Delta \Gamma
\simeq$0.08-0.10). There is an intrinsic correlation between the power-law spectral index and the
column density of a photoelectric absorber covering it as measured by the spectral fitting.
Steep spectra invariably correspond
to higher obscuration. In order to disentangle the intrinsic driver of the continuum spectral variability
we run two Scenario\#3 models, where the power-law index or the column density had been
constrained to assume alternatively the same value among all spectra. The quality of the fit - albeit
worse than when both parameters are allowed to vary - is better in the case when the power-law is allowed to vary:
$\chi^2/\nu = 1276.4$ against 1300.6/1164, suggesting that variations of the intrinsic continuum are
the main driver of this long-term continuum spectral variability.

\begin{figure}
  \includegraphics[height=80mm, angle=-90]{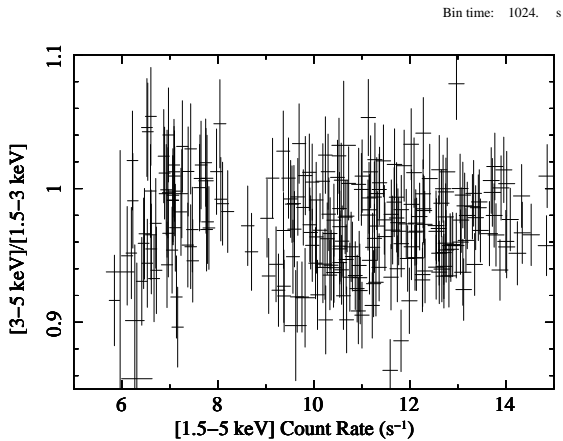}
  \caption{3--5~keV versus 1.5--3~keV hardness ratio as a function of the 1.5--5~keV count rate in all
  the XMM-Newton observations of Tab.~\ref{tab6}.}
  \label{fig7}
\end{figure}

\begin{figure}
  \includegraphics[height=80mm, angle=-90]{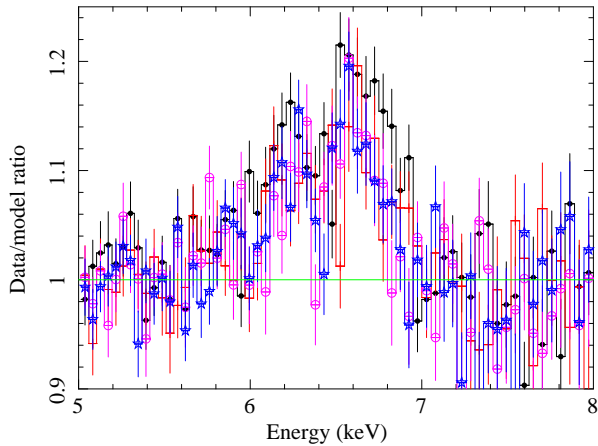}
  \caption{Residuals (in units of data/model ratio) when the best-fit continuum in Scenario\#3
is applied to intensity-resolved spectra in the nominal energy range excised of the 5--7~keV
interval. The contribution of the non-relativistic emission line components was
removed as well by subtracting their time-averaged intensity from the residuals. {\it Filled circles},
XMM78$_1$; {\it crosses}, XMM78$_2$; {\it empty circles}, XMM78$_3$; {\it stars}, XMM78$_4$.}
  \label{fig8}
\end{figure}

The X-ray light curve of NGC~5506 shows also rapid variability, by a factor $\simeq$1.7 over 250~s
($\Delta L_X$$\simeq$1.1$\times 10^{42}$~erg~s$^{-1}$; Dewangan \& Griffiths 2005).
The variability is achromatic (Fig.~\ref{fig7}; McHardy \& Czerny 1987).
In order to investigate the possible dependence of the broad line parameters upon the overall source
flux on such small timescales,
we have combined the event lists of the 2009 observations, and extracted four intensity-resolved spectra,
corresponding to the following background-subtracted count rate ranges: $<$2.9, 2.9--3.2, 3.2--3.5, and
$>$3.5~s$^{-1}$. These ranges were chosen in such a way that the intensity-resolved
spectra have approximately the same net background-subtracted counts.
These spectra are labelled as XMM78$_j$ in Tab.~\ref{tab4}, with $j$=1 to 4 (in increasing flux order).
The corresponding 2--10~keV absorption-corrected fluxes are:
$109.3 \pm^{1.3}_{1.4}$, $124.3 \pm^{1.3}_{1.6}$, $134.6 \pm^{1.6}_{1.8}$,
$145.0 \pm^{1.3}_{1.7}$$\times$10$^{-12}$~erg~cm$^{-2}$~s$^{-1}$, respectively.
No correlation exists between the EW of the broad K$_{\alpha}$
iron line component and the flux (Tab.~\ref{tab4}), save for a larger EW in the spectrum corresponding
to the lowest flux state.
The shape of the broad line component remains also remarkably constant in different flux
states (Fig.~\ref{fig8})

\section{Discussion}

\subsection{Continuum variability}

NGC~5506 is one of the most rapidly variable AGN of the X-ray sky. It exhibits luminosity changes of
$\sim$10$^{42}$~erg~s$^{-1}$ within a few minutes \cite{dewangan05}. The variability is, however,
remarkably achromatic on these time scales (Fig.~\ref{fig7}; see also
McHardy \& Czerny 1987). On longer timescales, a small but
systematic difference in the intrinsic spectral shape ($\Delta \Gamma \simeq 0.1$) is found
between ``low flux'' and
``high flux'' observations (absorption-corrected flux ratio between the latter and the former
of about 1.5). An even larger dynamical range in both luminosity and spectral shape
was measured during a hard X-rays
($E > 20$~keV) monitoring campaign with HEXIT \cite{manchanda06}.
The variability pattern measured by XMM-Newton is inconsistent with a pair-dominated corona \cite{haardt97},
where an increase of the spectral index by the amount observed by XMM-Newton
requires at least an order-of-magnitude
increase of the 2--10~keV luminosity.
This does not automatically imply that pairs are not an important source of scattering
in the Comptonizing material; the alternative possibility is that the increase of the corona
optical depth is due to a variation of the linear dimension of the corona. However, the shape
of the broad iron K$_{\alpha}$ iron line does not change significantly on the shortest
possible variability timescales which XMM-Newton can probe (Fig.~\ref{fig8}), suggesting
that the illumination of the disk does not significantly change.

\subsection{The origin of the narrow iron K$_{\alpha}$ line}

One of the main results presented in this paper is that
the intensity of the narrow component of the K$_{\alpha}$ in NGC~5506 remains constant during
the whole eight years covered by the XMM-Newton campaign. This extends by a factor of 3
the longest baseline published so far - a $\sim$1000~days RXTE monitoring campaign at the
end of last century \cite{lamer00}. The RXTE/PCA spectroscopic measurements were unable to
disentangle the two components (narrow and broad) of the iron K$_{\alpha}$ line. Nonetheless
Lamer et al. (2000) claimed that the strength of the continuum reflection
component and the total EW of the iron lines were stronger during phases of low flux.
The origin of the iron K$_{\alpha}$ emission line has challenged the ingenuity of observational
and theoretical astronomers since its early discovery \cite{mushotzky82,turner89}.
Evidence that this spectral feature is almost invariably accompanied by a Compton
reflection continuum component \cite{pounds90,nandra94} in unobscured AGN suggested
an origin in optically thick material out of the line of sight.
The origin of very broad and skewed profiles in the innermost region of the accretion disk
is a matter of heated debate, that we briefly summarised
in Sect.~1 (we will further discuss this point in
Sect~4.3). As far as narrower (typical Full Width Half Maximum, FWHM, $\approxlt$3000~km~s$^{-1}$)
and symmetric lines are concerned, possible physical
locations for the line-emitting matter are the
external regions of the accretion disk \cite{fabian89}, the Broad Line
Regions (BLRs) \cite{yaqoob01,bianchi08} or the molecular torus \cite{ghisellini94,krolik94}.
These systems have sizes from a fraction to several tens of parsecs.
The lack of correlation between the FWHM of the iron K$_{\alpha}$ and that of the
H$_{\beta}$ lines (or the black hole mass) disfavours a general origin in the BLRs \cite{nandra06}.

We can use the lack of variability of the narrow component of the iron K$_{\alpha}$ fluorescent line
to estimate the dimension of its production region.
The line produced by a reflector at distance $d$ from an illuminating source
tracks the variability pattern of the primary continuum if $d \approxlt t_{var}/c$,
where $t_{var}$ is the continuum variability timescale and $c$ is the speed of light.
We estimate $t_{var}$ as the longest time interval between XMM-Newton observations, whose
2--10~keV absorption-corrected fluxes differ by more than 27\%, {\it i.e.} than the
upper limit on the percentage fractional variability of the iron line intensity. $t_{var}$
hence corresponds to the span between observations XMM1 and XMM8: 7.92~years. This estimate
is qualitatively in agreement with the determination of the Power Spectrum Density
by the RXTE/PCA monitoring campaign \cite{uttley05}, which exhibits a slope of $\simeq$-1
below $\nu \simeq 4 \times 10^{-5}$~Hz and down to $\simeq$6$\times 10^{-9}$~Hz.
The XMM-Newton results constrain the location of the optically-thick
matter responsible for the bulk of the iron K$_{\alpha}$ narrow component at a distance
$\approxgt$2.5~pc. This is consistent with the best estimate of the inner side of the torus
($< 5$~pc; Prieto \& Meisenheimer 2004). Unfortunately no direct estimate of the Broad Line Region size
is available for NGC~5506. The 2--10~keV X-ray luminosity
from the data presented in this paper is in the range $\simeq$6--11$\times 10^{42}$~erg~s$^{-1}$, which
corresponds to a typical size of the BLR $\approxlt$10~days \cite{kaspi05}. This makes the torus the
most likely origin of the narrow iron K$_{\alpha}$ line.

We report a possible change in the intensity of the Fe{\sc xxvi} during two of the observations
in 2004 performed a few weeks apart, as well as a possible ionisation change of the
Fe{\sc xxv} (from a resonance- to a forbidden-dominated state) during XMM8.
Given, however, the spectral complexity
around 7~keV, a confirmation
of these findings through instrumentation with a much better energy resolution is needed. 
This is shown, for instance, by the different results on the forbidden component of the He-like
during XMM8 when one compares the phenomenological (Sect.~3.2) and the relativistic model (Tab.~\ref{tab2}).

\subsection{The origin of the broad iron K${\alpha}$ line}

This paper reports the discovery of a broad ($\sigma \simeq$330~eV) component of the iron K$_{\alpha}$
fluorescent emission line in NGC~5506. The line profile
lacks a prominent red wing, in contrast to what typically observed in unobscured Seyfert galaxies \cite{miller07}.
Although fits with a relativistically broadened profile yield marginally 
better statistical quality than symmetric Gaussian broadening, it is worth posing the question of
whether this difference may be due to a different physical mechanism
responsible for the observed broadening.

The symmetric and highly localised
profile rules out an explanation in terms of blending of transitions corresponding to
different iron ionisation states, enhanced continuum reflection \cite{guainazzi02,risaliti02},
scattering from a relativistic wind \cite{titarchuk09}
or incorrect fitting of the
opacity affecting the underlying continuum \cite{turner09}.
Comptonization of line photons was suggested to explain the first observations of broad lines by
{\it Ginga} \cite{czerny91}. The observed moderate width requires the line photons to
cross at most 3 Compton depths of
cool ($kT_C \le 0.7$~keV; $\sigma = \sqrt 2kT/mc^2$) gas \cite{pozdnyakov79,sunyaev80}.
The main shortcoming of
this scenario, however, is the non-linear dependence of the line EW
on the continuum flux on timescales as short as
a few hundreds seconds, which is at odds with the line-emitting region being screened by
optically thick matter.

The explanation in terms of a relativistically broadened and skewed profile is consistent
with the data.
Formal fits require a moderate inclination
($\imath \simeq$40$^{\circ}$) accretion disk.
NGC~5506 joins the so far restricted club of obscured AGN, which require relativistic
effects from a moderately inclined ($\imath \simeq 40^{\circ}$) accretion disk, together with
MCG-5-23-16 \cite{balestra04}
and NGC~526A \cite{nandra07}. All the above sources exhibit a column density obscuring the active nucleus
in the range 10$^{22-23}$~cm$^{-2}$. Although it is tempting to associate these sources with
AGN seen through the rim of the obscuring torus (see, for instance,
the discussion in Matt et al. 2003),
it should be borne in mind that
several unobscured AGN in the Nandra et al. (2007) also formally require large inclination
angles ({\it e.g.} Akn~120, Mkn~590, NGC~7213, NGC7469). On the other hand, the recently discovered
broad iron lines in IRAS13197-1627 \cite{dadina04,miniutti07} and
NGC~1365 \cite{risaliti09} require moderate disk inclinations of $\imath \simeq 27^{\circ}$ and 24$^{\circ}$,
respectively.
A comparison between unbiased samples is needed before
any inference on the geometrical configuration of the torus and the disk in X-ray obscured AGN can be drawn from
X-ray spectroscopy of the relativistically broadened K$_{\alpha}$ line.
Moreover, the disk inclination in NGC~5506 is significantly smaller than the inclination of the galaxy disk
\cite{imanishi00}.

The broad line is not extremely relativistic. It lacks the prominent red wing extending down to $\simeq$4~keV
observed in, {\it e.g.}, MCG-6-30-15 \cite{fabian03,miniutti07}, which requires
very steep radial emissivity profiles \cite{wilms01}. The inner radius of the line emitting region
(constrained by the data to be comprised between 3 and 25~$r_g$) and the flat emissivity profile suggest that a large area
on the disk contributes to the average profile. The line profile is remarkably constant in
intensity-resolved spectra, which sample variations of the intrinsic flux on time scales as low
as a few hundred seconds (cf. Fig.~\ref{fig8}) (against at odd with MCG-6-30-15; Iwasawa et al. 1999).
The fits do not require statistically significant changes in the structure of the innermost accretion
flow. The parameters $\alpha$ and $r_{in}$ describing the profile of the relativistically broadened
iron line are consistent with being constant over 5 epochs of observations spanning 9 years.

Little is known of
the black hole mass in NGC~5506. Both the central stellar velocity dispersion ($\simeq$180 km~s$^{-1}$)
\cite{oliva99,gebhardt00,ferrarese00,papadakis04}
and the width of the [OIII] line \cite{boroson03} suggest a black hole mass
$\sim$10$^8$M$_{\odot}$. For this mass the shortest measured X-ray variability timescale
implies that the bulk of high-energy continuum production and reprocessing occur within a few
Schwarzschild radii from the black hole. This is in disagreement with the lack of extreme relativistic line
broadening, as well as more generally with the idea that NLSy1 Galaxies are characterised by small black hole masses
and high accretion rates ($\approxlt$2\% in this scenario) \cite{komossa07}.
However, X-ray-based estimates suggest that the black hole mass in NGC~5506 could be significantly lower.
McHardy et al. (2006) observe that lowering the black hole mass by a factor $\sim$5 would bring NGC~5506
to the locus in the PDS break frequency, black hole mass and accretion rate plane shared by AGN and
Galactic Black Hole systems. An even more radical revision ($M_{BH} \simeq 5 \times 10^{6}$~M$_{\odot}$)
has been proposed by Niko\l ajuk et al. (2009) on the basis of the correlation with the
X-ray 2--10~keV light curve excess variance. A black hole mass of this order
($\sim$2$\times 10^6$~M$_{\odot}$) had been proposed by Hayashida et al. (1998) using similar arguments.
The light crossing time of the shortest variability
timescale if one assumes the Niko\l ajuk et al. (2009) estimate corresponds to about 20 Schwarzschild radii, in better
agreement with broad iron line spectroscopy results.
In this case the accretion rate is $\simeq$40\% of the Eddington limit, if one
estimates the bolometric luminosity by
applying a luminosity-dependent
bolometric correction to the 2--10~keV luminosity \cite{marconi04}. The size of the BLR,
$r_{BLR} \sim G M / v^2$, is
$\sim$8~days, if one estimates $v$ from the FWHM of the Pa$\beta$ broad component \cite{nagar02} .

The size of the AGN sample with robust detections of relativistic lines is still far too small for
correlations with other observables to yield stringent constraints. However, it is suggestive that many of the
objects of Nandra et al. (2007) ``dream team'' of relativistic AGN
are NLSy1s (Mkn~766, NGC~4051, NGC~5506, NGC~7314) or at least
characterised by extreme variability on short (MCG-6-30-15, NGC~4395) or long (NGC~2992) timescales.
Possible observational biases are discussed (and ruled out) by Nandra et al. (2007). 
A trend towards relativistically broadened lines being more common in Narrow Line AGN
seems to be present in the stacked spectra of large samples of unobscured AGN \cite{longinotti08}.
However, there doesn't seem to be a correlation between the detection of relativistically broadened iron
lines and either the accretion rate or the black hole mass in the, {\it e.g.}, FERO sample
\cite{longinotti08}.
Once again, the small size and incompleteness of the available samples hamper firm conclusions.

\section*{ACKNOWLEDGEMENTS}

Based on observations obtained with XMM-Newton, an ESA science mission with instruments and contributions
directly funded by ESA Member States and NASA. Useful discussions with Michal Dov\v ciak and
Giovanni Miniutti are gratefully acknowledged. A careful reading of the manuscript by an anonymous
referee significantly improved the presentation of the results.

\label{lastpage}

\end{document}